\documentclass[twocolumn,aps,prl,superscriptaddress]{revtex4-1}
\usepackage{amssymb} 
\usepackage{amsmath,bm} 

\usepackage{graphicx}
\usepackage{dcolumn}
\usepackage{bm}

\usepackage{multirow}
\usepackage[switch,columnwise]{lineno}

\usepackage[colorlinks=true,linktocpage=true,linkcolor=blue,citecolor=blue]{hyperref}

\graphicspath{{figures/}}

\begin{document}

\title{Probing the Neutron Skin with Extreme Collision Geometries in Heavy-Ion Collisions}

\author{Hui Zhang}
\address{State Key Laboratory of Nuclear Physics and Technology, Institute of Quantum Matter, South China Normal University, Guangzhou 510006, China.}
\affiliation{Guangdong Basic Research Center of Excellence for Structure and Fundamental Interactions of Matter, Guangdong Provincial Key Laboratory of Nuclear Science, Guangzhou 510006, China.}
\affiliation{Physics Department and Center for Exploration of Energy and Matter, Indiana University, 2401 N Milo B. Sampson Lane, Bloomington, IN 47408, USA.}

\author{Alex Akridge}
\address{Physics Department and Center for Exploration of Energy and Matter, Indiana University, 2401 N Milo B. Sampson Lane, Bloomington, IN 47408, USA.}

\author{Charles J. Horowitz}
\address{Physics Department and Center for Exploration of Energy and Matter, Indiana University, 2401 N Milo B. Sampson Lane, Bloomington, IN 47408, USA.}

\author{Jinfeng Liao}
\address{Physics Department and Center for Exploration of Energy and Matter, Indiana University, 2401 N Milo B. Sampson Lane, Bloomington, IN 47408, USA.}

\author{Hongxi Xing}\email{hxing@m.scnu.edu.cn}
\address{State Key Laboratory of Nuclear Physics and Technology, Institute of Quantum Matter, South China Normal University, Guangzhou 510006, China.}
\affiliation{Guangdong Basic Research Center of Excellence for Structure and Fundamental Interactions of Matter, Guangdong Provincial Key Laboratory of Nuclear Science, Guangzhou 510006, China.}
\affiliation{Southern Center for Nuclear-Science Theory (SCNT), Institute of Modern Physics, Chinese Academy of Sciences, Huizhou 516000, China.}

\date{\today}

\begin{abstract}

Understanding  how protons and neutrons are located differently in an atomic nucleus can provide fundamental information on nuclear structure and have far-reaching implications for astrophysics. A precise determination of this important difference, often quantified by the so-called neutron skin thickness, is challenging both theoretically and experimentally. Here we show how one can use a new category of observables in heavy ion collisions to probe the neutron skin thickness of nuclei like $^{208}$Pb and $^{48}$Ca, by utilizing the asymmetry between neutrons and protons of spectator nucleons in super-central collisions as well as that of participant nucleons in peripheral collisions. Using quantitative simulations, we demonstrate their sensitivity and great potential in constraining neutron skin thickness for both $^{208}$Pb and $^{48}$Ca nuclei in these extreme event geometries. Furthermore, we propose the asymmetric collisions between $^{48}$Ca and $^{40}$Ca nuclei as a unique and powerful way to nail down the neutron skin thickness.      

\end{abstract}

\maketitle

\section{Introduction}

An important layer of the hierarchical structure for matter in our physical universe consists of various nuclei that are at the heart of atoms. Describing the nuclear structure in terms of its even smaller constituents, namely the protons and neutrons, is a challenging problem at the forefront of nuclear physics~\cite{osti_2280968,osti_1999724,Nazarewicz:2016gyu,Bogner:2009bt}. Remarkably, microscopic dynamics governing nuclear structure also directly influences the properties of neutron stars in the sky and is essential for explaining today's multi-messenger observations in astrophysics and cosmology~\cite{Fattoyev:2017jql,Horowitz:2019piw,Piekarewicz:2019ahf,Wei:2019mdj,Steiner:2011ft,Gandolfi:2013baa,Li:2021thg}.   Despite significant progress in both theory and experiment,  some key questions regarding how protons and neutrons are distributed in a nucleus remain to be fully understood. An outstanding example is the neutron skin thickness~\cite{Abrahamyan:2012gp,PREX:2021umo,CREX:2022kgg,Thiel:2019tkm,Reed:2021nqk,Tarbert:2013jze}, which quantifies how differently the protons and neutrons are located. Therefore, developing new ways of constraining this quantity with laboratory experiments could have far-reaching impacts. 

While the neutron skin thickness is primarily constrained by electron scattering experiments at low energies \cite{Fricke:1995zz,Horowitz:1999fk,Abrahamyan:2012gp,PREX:2021umo,CREX:2022kgg}, measurements from relativistic heavy ion collisions have emerged as a promising, unconventional tool with strong potential for studying various aspects of nuclear structure~\cite{STAR:2024wgy,STAR:2025vbp,Jia:2022ozr,Jia:2021tzt,Liu:2023pav,Giacalone:2019pca,Giacalone:2021udy,Giacalone:2021uhj,Giacalone:2023cet,Liu:2022xlm,Li:2019kkh,Dobaczewski:2025rdi,Bally:2021qys,Ryssens:2023fkv,Zhang:2021kxj,Giacalone:2024luz,Giacalone:2024ixe,Fortier:2024yxs,Zhang:2025raf, Paukkunen:2015bwa, Pihan:2025pep, vanderSchee:2023uii}. Most of these studies utilize the collective flow observables that directly probe the initial conditions of the collisions, which in turn depend on the geometric deformations in the nuclear structure inputs. It has been demonstrated that certain geometric features in the size and shape of the colliding nuclei could manifest themselves in the final-state hadrons' flow or flow correlations. In this work, we propose a new category of observables that are not based on collective flow but rather leverage the asymmetry between neutrons and protons of the participant as well as spectator nucleons in these collisions. As we shall demonstrate in the rest of this paper, such observables can help provide important new constraints on the neutron skin thickness of nuclei like $^{208}$Pb and $^{48}$Ca.

\section{Accessing Neutron Skin in Heavy Ion Collisions}

The density distributions of protons and neutrons inside an atomic nucleus with $Z$ number of protons and $(A-Z)$ number of neutrons can be well parametrized by the so-called Woods-Saxon distributions: 
\begin{eqnarray} \label{eq_rho}
    \rho_i(r)=  \frac{\rho_{i 0}}{1+e^{(r-R_i)/a_i}}, 
\end{eqnarray}
where $i=n,p$ for neutrons and protons, respectively. While angular dependence can be introduced in the above distributions to describe shape deformation, here we focus on 
the neutron-rich doubly-closed-shell (proton and neutron) nuclei, as is the case for $^{208}$Pb and $^{48}$Ca. 
For given radius parameter $R_i$ and surface parameter $a_i$, the corresponding central density $\rho_{i 0}$ can be determined from normalization conditions: $\int d^3\vec{r} \rho_p(r) = Z$ and $\int d^3\vec{r} \rho_n(r) = (A-Z)$.  

Based on the distributions in Eq.~(\ref{eq_rho}), one can define the following mean-squared-radius for the neutrons and protons respectively:  
\begin{eqnarray}
    \langle r^2 \rangle_i = \frac{\int d^3\vec{r}\ r^2\ \rho_i(r)}{\int d^3\vec{r}\ \rho_i(r)}.
\end{eqnarray}
The difference between the neutron and proton distributions can then be conveniently quantified by the so-called neutron skin thickness, defined as:
\begin{eqnarray}
    \delta_{np}= \sqrt{\langle r^2 \rangle}_n -  \sqrt{\langle r^2 \rangle}_p.
\end{eqnarray}
A positive $\delta_{np}$ implies a wider spread of neutrons toward the edge of the nucleus compared to protons, which could be due to a larger $R_n$ or $a_n$ or both. 

The proton distribution parameters $R_p$ and $a_p$ can be well constrained by experiments like elastic electron scattering experiments. The neutron parameters $R_n$ and $a_n$ are much harder to measure. The best constraints so far come from parity-violating electron scattering experiments such as the PREX~\cite{PREX:2021umo}  reporting  $\delta_{np} = 0.283 \pm 0.071\ \rm fm$ for $^{208}$Pb and CREX~\cite{CREX:2022kgg}  reporting $\delta_{np} = 0.121 \pm 0.035\ \rm fm$ for $^{48}$Ca. Nuclear structure models are typically calibrated to describe the proton distribution parameters well, while their predictions for the neutron distribution parameters (especially for $R_n$) often differ considerably~\cite{Horowitz:2020evx,Reed:2020fdf}.

To see how the neutron skin may manifest itself in heavy ion collisions, let us use $^{208}$Pb as an example. With a positive $\delta_{np}$, its surface then has a layer of ``skin'' consisting predominantly of neutrons. In a super-central collision where the two incident nuclei overlap nearly perfectly, the non-overlapping spectator region coincides with side patches of the two nuclei' neutron skins. Clearly, the larger $\delta_{np}$ is, the more prominent this spectator effect would be.  
These spectator neutrons from the skin patches travel down the beamline after collisions and can be measured by the so-called Zero-Degree Calorimeter (ZDC) detectors installed at both ends of the beamline. Charged particles like spectator protons are diverted away by the dipole magnets down the beamline and by design will not reach the ZDCs. The ALICE, ATLAS and CMS collaborations at LHC as well as the STAR collaboration at RHIC have all demonstrated excellent capability of utilizing the ZDC detectors to capture and count spectator neutrons; see details in e.g. ~\cite{ALICE:2022iqi,ATLAS:2020epq,CMS:2020skx,Xu:2016alq}.  
Therefore, the number of spectator neutrons in such super-central collisions offers a direct and sensitive probe to neutron skin thickness. We emphasize that the key point here is to focus on spectator neutrons in super-central collisions (e.g. in the $0\sim 1\%$ or even $0\sim 0.1\%$ centrality), where the neutron skin effect is maximized and where there are typically just few spectator neutrons for the benefit of optimal ZDC counting. 

In very peripheral collisions, on the other hand, the two nuclei overlap only marginally, and it is the participant region that coincides with patches of the neutron skins. As a result, the participant nucleons in such collisions will be ``neutron rich'', i.e. containing a higher fraction of neutrons than the baseline expectation of (A-Z)/A. The relativistic collisions of these participant nucleons would then create a quark-gluon plasma with a higher net isospoin in initial conditions  compared with a scenario without neutron skin. The final-state hadrons produced from the QGP, dominantly pions,  will inherit such isospin initial condition that can  be quantified by yield difference between $\pi^-$ and $\pi^+$ hadrons.  

In the following, we use quantitative simulations to predict the magnitude of these two observables from the neutron skin. We emphasize that the proposed method here is new and different from previous approaches of accessing geometric features in nuclear structure via flow observables.  
While $^{208}$Pb is used as a primary example of our discussion, the same idea can certainly be applied to $^{48}$Ca and other nuclei alike.

\section{Spectator Neutrons in Super-Central Collisions} 

To study the impact of neutron skin thickness on initial conditions, we use the Monte Carlo (MC) Glauber model  which has been widely used and data-validated for simulations of heavy ion collisions~\cite{Miller:2007ri,Loizides:2017ack}. Here we focus on PbPb collisions at LHC energy $\sqrt{s_{NN}}=5.02~\rm TeV$. The nuclear structure inputs for Glauber model come from the nucleon distributions in Eq.~\ref{eq_rho}, with protons and neutrons in the nucleus being sampled according to their respective $\rho_i$ on an event-by-event basis.  We adopt the commonly accepted parameters $R_p=6.70\ \rm fm$ and $a_p=0.447\ \rm fm$ for proton distributions. For neutron distributions,  density functional models suggest a well-informed $a_n\approx 0.566\ \rm fm$~\footnote{ Using results from fourteen relativistic mean field  models, one obtains an estimate of the mean and variance to be: $a_n\approx (0.566 \pm 0.009) \ \rm fm$. (Based on \cite{Reed:2020fdf} and  private communications with Brandon Reed.)}.  It may be noted that both $a_n$ and $a_p$ are constrained by the known surface energy term in the semi-empirical mass formula.  In this study, we consider the neutron distribution radius $R_n$ as a tunable parameter~\cite{Horowitz:2020evx,Reed:2020fdf}. 
For fixed values of $R_p,a_p$ and $a_n$, varying the $R_n$ is equivalent to varying the neutron skin thickness $\delta_{np}$. In the following, we will simply use $\delta_{np}$ as our ``control parameter''.  

For each given set of parameters, we perform MC Glauber simulations with a total of 10 million events. One can then examine the numbers of participant protons, participant neutrons, spectator protons, and spectator neutrons respectively in different centrality class. We observe two clear trends with varying neutron skin thickness: with an increasing $\delta_{np}$, the spectator neutrons relative to spectator protons increase rapidly in very central collisions, while the participant neutrons relative to participant protons increase rapidly in peripheral collisions. Such quantitative results confirm the qualitative insights discussed above.   

\begin{figure}[!htb]
    \includegraphics[width=200pt]{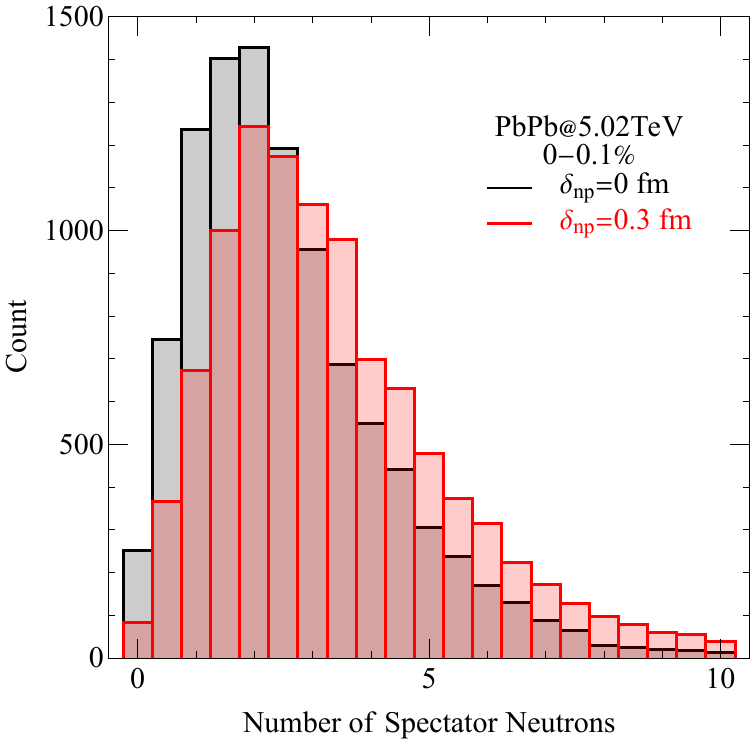}
    \caption{Histogram for the number of spectator neutrons in super-central ($ 0\sim0.1\%$) PbPb collisions at $\sqrt{s_{NN}}=5.02~\rm TeV$ from simulations with thickness $\delta_{np}=0.3\ \rm fm$ (red) and $\delta_{np}=0\ \rm fm$ (black), respectively.   }
    \label{fig:Neutron1}
\end{figure}

\begin{figure}[!htb]
    \includegraphics[width=200pt]{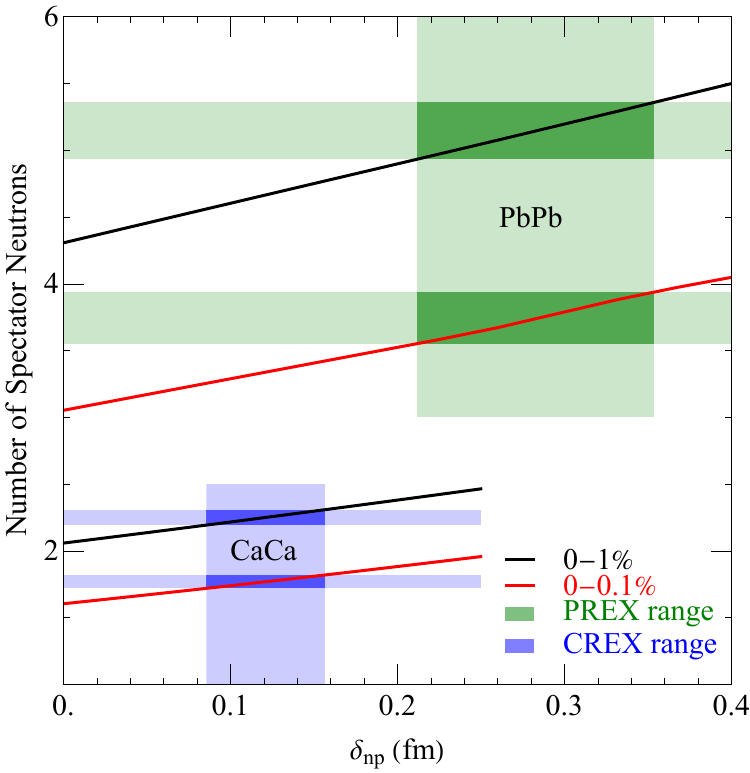}  
    \caption{The average number of spectator neutrons (either forward or backward) versus neutron skin thickness $\delta_{np}$ in super-central PbPb and CaCa collisions at $\sqrt{s_{NN}}=5.02~\rm TeV$, with red curve for $ 0\sim0.1\%$, black curve for $ 0\sim 1\%$.  The green (blue) shaded bands indicate projected observable range based on current PREX (CREX) results for $\delta_{np}$ of $^{208}$Pb ($^{48}$Ca), respectively.    }
    \label{fig:Neutron2}
\end{figure}

To best reveal the neutron skin effect, we propose to count spectator neutrons in super-central collisions with ZDC detectors. This is illustrated for the $0\sim 0.1\%$ collisions in Fig.~\ref{fig:Neutron1}, where a thickness $\delta_{np}=0.3\ \rm fm$ leads to a considerable increase in the number of spectator neutrons as compared with $\delta_{np}=0\ \rm fm$ case. 
One can use the average number of spectator neutrons (either forward or backward) as a concrete observable, which is shown as a function of thickness $\delta_{np}$ in Fig.~\ref{fig:Neutron2} for both $0\sim 0.1\%$ and $0\sim 1\%$ centrality bins. The green shaded bands indicate the projected observable range (on the vertical axis) based upon current PREX results for $\delta_{np}$, with an uncertainty at about $25\%$ level. The same observable has also been studied for $^{48}$Ca$^{48}$Ca collisions, with results also shown in Fig.~\ref{fig:Neutron2}. For $^{48}$Ca calculations, we use $R_p= 3.74\ \rm fm$, $a_p=a_n= 0.525\ \rm fm$, and vary $R_n$. 
These results suggest a future precision measurement of this observable has the potential of significantly improving the constraint on neutron skin thickness of both  $^{208}$Pb and $^{48}$Ca. Considering practical measurements, we note that the $0\sim 0.1\%$ centrality case with fewer neutrons offers good feasibility for current ZDC detectors. On the other hand, the  $0\sim 1\%$ centrality case could be more statistics friendly.

\section{Pion Yield Asymmetry in Peripheral Collisions}

We now turn to the peripheral collisions, in which the fraction of neutrons relative to protons in the participants is boosted by an increasing neutron skin thickness. Suppose there are $N_n$ and $N_p$ participant neutrons and protons, respectively, one can quantify the difference with a ratio $(N_n-N_p)/(N_n+N_p)$. This asymmetry between participant neutrons and participant protons is inherited by the quark-gluon plasma formed after the initial collisions, in which the initial net quark densities $n_u$ and $n_d$ for the u and d flavors should satisfy $(n_d-n_u)/(n_d+n_u) \approx (1/3) \cdot (N_n-N_p)/(N_n+N_p)$. Such an initial condition will eventually translate into the yields in final-state hadrons after the system's evolution. Specifically, we construct the following observable $R \equiv (Y_{\pi^-} - Y_{\pi^+})/ (Y_p - Y_{\bar p})$, where $Y_{\pi^-}, Y_{\pi^+}, Y_{p}, Y_{\bar p}$ are the yields of $\pi^-$, $\pi^+$, proton and anti-proton, respectively. In this ratio, the numerator and denominator serve as proxies for isospin and net baryon number, respectively.

\begin{figure}[!htb]
    \includegraphics[width=200pt]{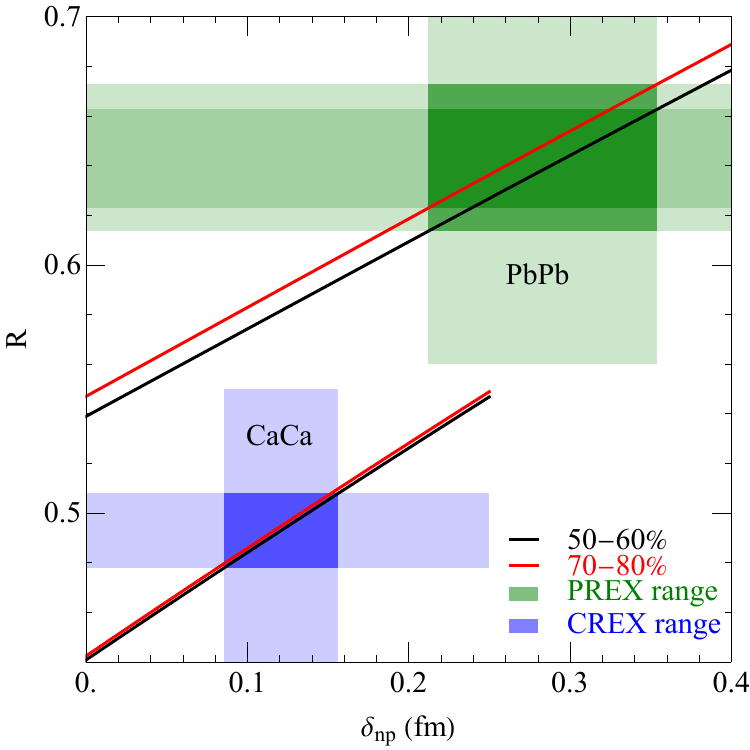}
    \caption{ The yield asymmetry    observable $R$ (as defined in text) versus neutron skin thickness $\delta_{np}$ in peripheral PbPb and CaCa collisions at $\sqrt{s_{NN}}=5.02 ~\rm TeV$, with red curve for $ 70\sim 80 \%$, black curve for $ 50\sim 60\%$.  The green (blue) shaded bands indicate projected observable range based on current PREX (CREX) results for $\delta_{np}$ of $^{208}$Pb ($^{48}$Ca), respectively.  }
    \label{fig:isospin}
\end{figure}

We next use the AVFD hydrodynamic simulation framework~\cite{Jiang:2016wve,Shi:2017cpu,Shi:2019wzi}  to quantify the response of this observable to the initial neutron/proton asymmetry. Specifically,  small initial net quark densities $n_u$ and $n_d$ in AVFD are set to satisfy $(n_d-n_u)/(n_d+n_u) = (1/3) \cdot (N_n-N_p)/(N_n+N_p)$ where the participant numbers $N_n$ and $N_p$ are computed from MC Glauber for given value of $\delta_{np}$. The results for both PbPb and CaCa collisions are shown in Fig.~\ref{fig:isospin}. As one can see, the observable $R$ increases linearly with $\delta_{np}$.  The green (blue) shaded bands indicate the projected observable range (on the vertical axis) based upon current PREX (CREX) results for $\delta_{np}$ of $^{208}$Pb and  $^{48}$Ca, respectively. Again, an acuate future measurement of   $R$ in  heavy ion collisions could help put an independent new constraint on the neutron skin thickness.

\section{Asymmetric Collisions of $^{40}$Ca$^{48}$Ca} 

The neutron skin thickness of $^{48}$Ca nucleus is of great interest. Here we propose another unique opportunity to constrain $\delta_{np}$ of $^{48}$Ca, by utilizing its     isotope nucleus $^{40}$Ca. Both $^{40}$Ca (with 20 protons and 20 neutrons) and $^{48}$Ca (with 20 protons and 28 neutrons) belong to the so-called ``double magic'' nuclei, with 8 extra neutrons leading to a sizable neutron skin for the latter~\footnote{For simplicity,  
we assume $\delta_{np}=0$ for $^{40}$Ca. This
neglects a rather small negative $\delta_{np}\approx -0.05 \ \rm fm $ for $^{40}$Ca due to the Coulomb effect among protons, which also  applies to protons in $^{48}$Ca. See: C.~J.~Horowitz and B.~D.~Serot,
``Self-consistent Hartree Description of Finite Nuclei in a Relativistic Quantum Field Theory,''
Nucl. Phys. A \textbf{368}, 503-528 (1981).}.  
Our proposal is to collide the $^{40}$Ca nucleus with the $^{48}$Ca nucleus and examine the spectator neutrons in super-central collisions where the centers of the two nuclei coincide almost perfectly. In this setup, the edge of $^{48}$Ca, dominated by its neutron skin, leads to a larger number of spectator neutrons in the  $^{48}$Ca-going direction (which we refer to as ``forward'') than that in the $^{40}$Ca-going direction (which we refer to as ``backward''). This asymmetry between $N_F$ (forward spectator neutrons) and $N_B$ (backward spectator neutrons), which can be detected by ZDCs, should provide a sensitive probe to the $\delta_{np}$ of $^{48}$Ca nucleus.

\begin{figure}[htb!]
    \includegraphics[width=200pt]{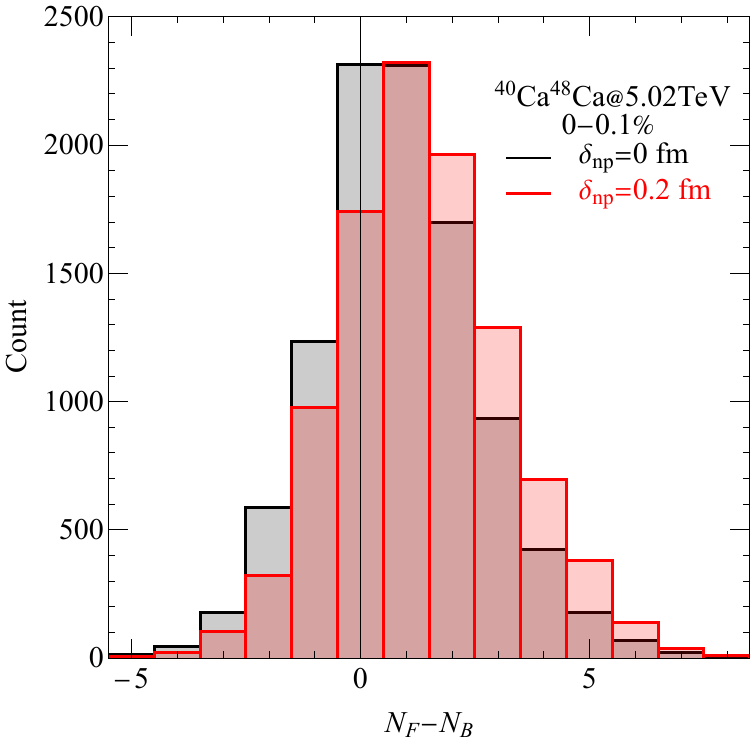} 
    \caption{Histogram of event-by-event difference between forward ($^{48}$Ca-going direction) and backward ($^{40}$Ca-going direction) spectator neutrons in $0\sim 0.1 \%$ centrality $^{40}$Ca$^{48}$Ca collisions at $\sqrt{s_{NN}}=5.02~\rm TeV$ from simulations with thickness $\delta_{np}=0.2\ \rm fm$ (red) and $\delta_{np}=0\ \rm fm$ (black), respectively.  }
    \label{fig:ca01}
\end{figure}

\begin{figure}[htb!] 
       \includegraphics[width=200pt]{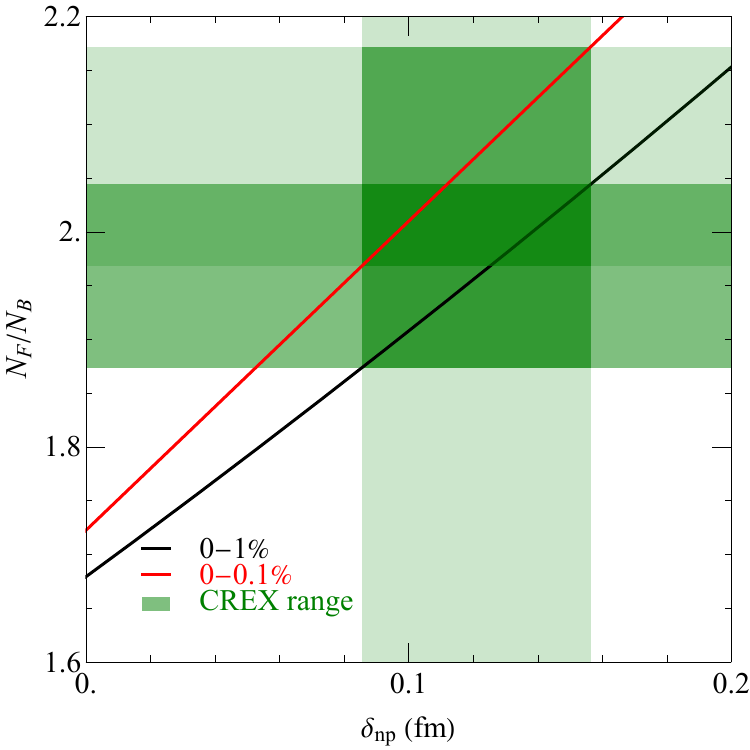} 
    \caption{ The ratio between event-averaged forward and backward spectator neutrons versus neutron skin thickness  $\delta_{np}$ in  $^{40}$Ca$^{48}$Ca collisions at $\sqrt{s_{NN}}=5.02~\rm TeV$, with red curve for $ 0\sim 0.1 \%$, black curve for $ 0\sim 1\%$, and green shaded bands for projected range based on current CREX results for $\delta_{np}$.  }
    \label{fig:ca02}
\end{figure}

We use Monte Carlo Glauber simulations to demonstrate this effect in $^{40}$Ca$^{48}$Ca collisions. Fig.~\ref{fig:ca01} shows the histogram of event-by-event difference between forward and backward spectator neutrons for $0\sim 0.1 \%$ centrality. The $\delta_{np}=0.2\ \rm fm$ case with a large neutron skin leads to a visibly larger forward-backward asymmetry in comparison with the $\delta_{np}=0\ \rm fm$ case. A new observable, the ratio between event-averaged forward and backward spectator neutrons, can be introduced to quantitatively probe the value of $\delta_{np}$. Their relation is shown in Fig.~\ref{fig:ca02} for both $0\sim 0.1 \%$ and $0\sim 1 \%$ centrality. The green shaded bands indicate the projected observable range (on the vertical axis) based upon current CREX results for $\delta_{np}$, which suggest a significant asymmetry (roughly a factor of 2) with an uncertainty level close to $30 \%$. This leaves considerable room for an improved constraint on $^{48}$Ca neutron skin thickness from heavy ion measurements and therefore provides strong motivation for carrying out asymmetric $^{40}$Ca$^{48}$Ca collisions in the future.

\section{Conclusion}
 
To summarize, we have proposed a new category of observables in heavy ion collisions to help constrain the neutron skin thickness of nuclei like $^{208}$Pb and $^{48}$Ca. These include the number of spectator neutrons in super-central collisions and the neutron/proton asymmetry of participants in peripheral collisions. The former can be directly counted by utilizing ZDC detectors at RHIC and LHC while the latter can be measured via the final-state hadron yield ratio observable $R = (Y_{\pi^-} - Y_{\pi^+})/ (Y_p - Y_{\bar p})$. We have used quantitative simulations to show their excellent sensitivity as probes of neutron skin thickness $\delta_{np}$ for both $^{208}$Pb$^{208}$Pb and $^{48}$Ca$^{48}$Ca collisions. We've further proposed and demonstrated that the asymmetry between forward and backward spectator neutrons in $^{40}$Ca$^{48}$Ca collisions provides a unique and powerful way of constraining  $^{48}$Ca neutron skin thickness. 
All taken together, we conclude that future measurements of these novel observables can offer exciting new opportunity to provide an independent experimental probe of the neutron skin thickness and have the potential of substantially improving current constraints on this fundamental knowledge about the structures of atomic nucleus.  

This study focuses on $^{208}$Pb and $^{48}$Ca, which are the only stable neutron-rich doubly-closed-shell nuclei. 
Radioactive beams would allow the study of other interesting examples of neutron-rich doubly-closed-shell nuclei such as the $^{132}$Sn, which is intermediate in size between $^{208}$Pb and $^{48}$Ca.  Colliding a nucleus like $^{132}$Sn and measuring the proposed observables of the present work may  shed light on the apparent thick skin in $^{208}$Pb and thin skin in $^{48}$Ca.

\vspace{0.2in}

\section*{Acknowledgments}

The authors are grateful to Brendan Reed and Shuai Yang for very helpful communications and discussions. HZ and HX acknowledges support from the National Natural Science Foundation of China (Grant Nos. 12525508, 12475139).  AA and JL acknowledge support by the U.S. NSF under Grant No.~PHY-2514992.  CH acknowledges support from US DOE grant DE-FG02-87ER40365 and US NSF grant PHY-2116686. 



\end{document}